\documentclass[useAMs,usenatbib,a4paper]{mn2e}
\usepackage{savesym}
\usepackage{graphicx}
\expandafter\let\csname equation*\endcsname\relax
  \expandafter\let\csname endequation*\endcsname\relax 
\usepackage{subfig}
\usepackage{amsmath}
\usepackage{verbatim}
\usepackage{array}
\usepackage[total={17.8cm,24.0cm},centering]{geometry}
\usepackage{times}

\newcommand{\be}{\begin{equation}}
\newcommand{\ee}{\end{equation}}
\newcommand{\eea}{\end{eqnarray}}
\newcommand{\bea}{\begin{eqnarray}}

\newcommand{\m}{\mathrm}
\newcommand{\gray}{$\gamma$-ray\,}

\title[Synchrotron and inverse-Compton from Conical Jets]{Synchrotron and inverse-Compton emission from blazar jets I: a uniform conical jet model.}
\author[William J. Potter and Garret Cotter]{William J. Potter\thanks{E-mail:
will.potter@astro.ox.ac.uk (WJP)} and Garret Cotter
\\
Oxford Astrophysics. Denys Wilkinson Building, Keble Road, Oxford, OX1 3RH, United Kingdom}
\begin{document}

\date{\today}

\maketitle

\label{firstpage}

\begin{abstract}
In the first of a series of papers investigating emission from blazar jets from radio to high-energy gamma-rays, we revisit the class of models where the jet has a uniform conical ballistic structure. We argue that by using simple developments of these models,  in the context of new multi-frequency data extending to \gray energies, valuable insights may be obtained into the properties that fully realistic models must ultimately have. In this paper we consider the synchrotron and synchrotron-self-Compton emission from the jet, modelling the recent simultaneous multi-wavelength observations of BL Lac. This is the first time these components  have been fitted simultaneously for a blazar using a conical jet model. 

In the model we evolve the electron population dynamically along the jet taking into account the synchrotron and inverse-Compton losses. The inverse-Compton emission is calculated using the Klein-Nishina cross section and a relativistic transformation into the jet frame, and we explicitly show the seed photon population.  We integrate synchrotron opacity along the line of sight through the jet plasma, taking into account the emission and opacity of each section of the jet. In agreement with previous studies of radio emission, we find that a conical jet model which conserves magnetic energy produces the characteristic blazar flat radio spectrum, however, we do not require any fine-tuning of the model to achieve this. Of particular note, in our model fit to  BL Lac---which at $\sim 10^{37} W$ is a relatively low jet-power source---we find no requirement for significant re-acceleration within the jet to explain the observed spectrum. 

\end{abstract}


\section{Introduction}

Blazars are the most luminous and highly variable of active galactic nuclei (AGN).  They have been observed at all wavelengths from radio to high energy $\gamma$-rays and are believed to be AGN observed at small angles to the jet axis (\cite{1995PASP..107..803U}) resulting in Doppler boosting of their emission.  The characteristic spectral shape is a non-thermal continuum with two main peaked components; the radio through to UV/x-rays show signs of polarisation typical of synchrotron radiation, the second component extends from low energy x-rays through to very high energy $\gamma$-rays and is normally attributed to inverse-Compton scattering of photons from the synchrotron-emitting electrons.  The radio emission of blazars is predominantly observed to be flat or nearly flat in flux density (see for example \cite{1980AJ.....85..351O} and \cite{2010ApJ...716...30A}) whilst the majority of the synchrotron and inverse-Compton emission can be well described by a power law and a turnover.

In general, VLBI images of AGN observed at large angles to the jet axis show a continuous axisymmetric plasma jet, which is approximately conical in geometry, (see e.g., \cite{2007ApJ...668L..27K} and \cite{Krichbaum:2006pw}) with evidence for a small degree of curvature close to the base of the jet (\cite{2011arXiv1110.1793A}) and a blunt base with a wide opening angle (\cite{2011Natur.477..185H}).  Recent work by \cite{2011arXiv1103.6032S} measured the frequency dependent core-shift for 20 AGN and found the results favoured a conical jet which conserved magnetic energy.  

The early and influential model of \cite{1979ApJ...232...34B} used a uniform conical structure for the jet, and was able to produce the flat radio spectrum if magnetic energy in the jet was conserved. Further investigations by \cite{1980ApJ...235..386M}, \cite{1981ApJ...243..700K}, \cite{1982ApJ...256...13R}, \cite{1984ApJ...285..571M} and \cite{1985A&A...146..204G} indicated that this class of model was a good fit to the observed synchrotron emission. However, at the time, \gray observations were rare, and replenishment of adiabatic and radiative losses in some of the models were deemed to be problematic.  

More recently, models for the radio and inverse-Compton emission from relativistic jets have predominantly considered all the emission to come from a small region of relativistic plasma, and such models have the ability to match the \gray spectrum of flares in blazars, e.g. \cite{1998A&A...333..452K}, \cite{2002ApJ...581..127B}, \cite{2000ApJ...536..729L} and \cite{2007A&A...476.1151T}. The spherical blob is injected with a power-law (or broken power-law) electron energy spectrum at regular time intervals and the distribution of electron energies is evolved using the continuity or Fokker-Planck equation and the resulting emission spectrum is averaged over time.  The continuity/Fokker-Planck equation takes into account the change in the electron population due to electron energy losses from emission and due to the injection of additional electrons, but the energy injection mechanism remains arbitrary.

Models which contain some extended jet component explicitly include those of, e.g., \cite{1985ApJ...298..114M}, \cite{2000A&A...356..975K}, \cite{2000AAS...197.8401M}, \cite{2001MNRAS.325.1559S} and \cite{2010MNRAS.401..394J}. These have variously included some attempts to treat particle re-acceleration, but are often aimed at x-ray binary systems and have not been concerned with the entire spectrum of a blazar from radio to \gray. Recently the conical jet was investigated by \cite{2006MNRAS.367.1083K} as part of an analytic  study of the range of conditions in the jet plasma which could result in a flat radio spectrum, although this model too did not investigate the inverse-Compton component of the spectrum. 

With the advent of high-quality simultaneous radio-to-\gray spectra, especially including data from the {\sl Fermi} satellite, we were motivated to re-visit the problem of the full blazar spectrum.  In particular, we wish to stress 
that the ultimate goal should be a model which contains both explicit and realistic jet parameters along with a detailed treatment of the particle emission processes and energetics within the jet.  

In this paper we set out the details of a flexible and rigorous model which will form the basis of a series of investigations into the properties of blazars.  We use a simple conical jet geometry as a first-order model for the structure of the jet. Our focus is on accurately calculating both the synchrotron and the synchrotron-self-Compton components of the jet spectrum. We do this using a full Klein-Nishina calculation of the inverse-Compton scattering and integrating the synchrotron emission along the line of sight through the relativistic jet, and we compare with the recent observations of BL Lac from radio to \gray. For a conical jet this is the first time a full-spectrum fit to a blazar has been attempted, and we also recover the jet opening angle, equipartition fraction and bulk kinetic power. 

In this first paper we investigate whether a simple ballistic conical jet can reproduce the spectrum of BL Lac without need for significant acceleration or external Compton radiation.   We include radiative electron losses but we do not yet include adiabatic losses in the jet. For the relatively low-power sources such as this, we find that there is not a need for significant re-acceleration within the jet.  In the next paper in the series we analyse whether this is the case for Compton-dominant high energy sources.  

Most previous investigations of the conical jet have used simplifying approximations when calculating the opacity and observed synchrotron emission at small angles through the jet and approximate analytical expressions for the inverse-Compton emission.  We treat the calculations of the line of sight synchrotron opacity and inverse-Compton emission thoroughly.  We use numerical integration of exact expressions to calculate the inverse-Compton emission.  Unlike all previous investigations we also take into account electron energy losses along the jet due to both synchrotron and inverse-Compton emission and take into account the effect of these energy losses on the line of sight synchrotron opacity.       

In this paper we will first introduce the parameters and assumptions of our jet model including its geometry and electron population.  We then calculate the inverse-Compton spectrum produced from the scattering of an arbitrary isotropic seed photon population, although we shall only consider SSC seed photons in this work.  We calculate the observed synchrotron emission from the jet and the synchrotron photon population.  We show that the ballistic conical jet model is consistent with recent multi-frequency observations of BL Lacertae.  We show that this model reproduces the observed flat radio spectrum if the magnetic energy is conserved along the jet and that this flat radio spectrum is a property which is insensitive to the model's parameters.  Finally, we explicitly show the seed photon distribution and the evolution of the electron energy distribution.

\section{Jet Model}

\begin{figure}
\centering
\includegraphics[width=8cm]{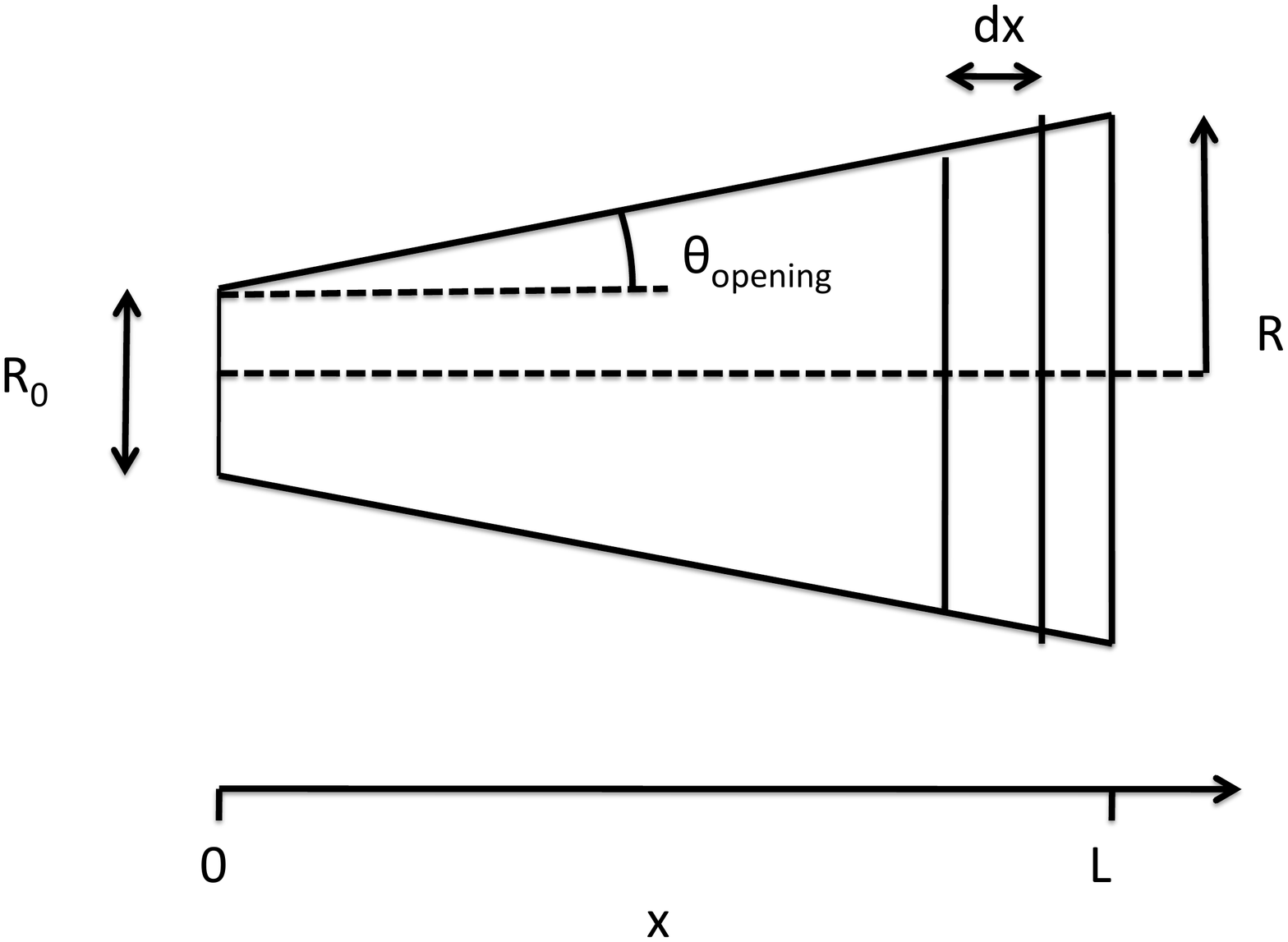}
\caption{Schematic diagram of the conical jet, showing parameters defined in the fluid centre of momentum frame}
\label{fig:jetdiag}
\end{figure}

As described above, VLBA images of AGN indicate that jets flows are axisymmetric with a constant opening angle over much of the jet length and so can be modelled by a conical geometry.  This geometry was also favoured by \cite{2011arXiv1103.6032S} to explain the frequency dependent core-shifts they measured.  We will assume that the jet has an opening angle $\theta_{\mathrm{opening}}$, initial radius $R_{0}$ and length $L$, all defined in the centre of momentum frame of the fluid (fluid frame), see Figure $\ref{fig:jetdiag}$.  We assume that the jet is composed of an electron-positron plasma (for simplicity we will use `electrons' to refer to both the electrons and positrons) which is injected at the base of the jet and the electrons lose energy radiatively as they move along the jet.  Further, we make the assumption that both the electrons and magnetic field are isotropically and homogeneously distributed within the fluid rest frame whilst the fluid is moving in the lab frame with an associated bulk Lorentz factor $\gamma_{\mathrm{bulk}}$.

We assume that the jet structure is constant with time in the lab frame and that the geometry of the jet is that of a truncated cone of length $\gamma_{\m{bulk}}L$ where the length of the jet in the lab frame is related to that in the fluid frame by a simple Lorentz contraction.  We define the variable $x$ as the length along the jet axis (in the fluid frame), where $x=0$ is defined dynamically as the base of the jet and $x=L$ as the end of the jet in the fluid frame.  The definition of $x$ is dynamic because in the fluid frame the structure of the jet is moving whilst the plasma is stationary, as opposed to a moving plasma and stationary jet structure in the lab frame (simply due to the relative velocity of the plasma and jet structure).  We can now parameterise all of our physical quantities as functions of $x$.  We also define the radius of the jet as $R(x)$ where

\be
R(x) = R_{0} + x \tan \theta_{\mathrm{opening}} .
\ee

\subsection{Determining the Jet Parameters at the Base}

We consider a jet with total power in the lab frame $W_{j}$.  At the base of the jet the jet material has speed $\beta c$ (where we assume a relativistic flow $\beta \simeq 1$).  So the energy contained in a section of plasma of width one metre in the $x$-direction in the lab frame is $E_{j}$ which is equal to $\frac{W_{j}}{c}$.  If the jet has a bulk Lorentz factor $\gamma_{\mathrm{bulk}}$ then the energy $E_{j}$ in the lab frame is related to the energy contained in a one metre section of plasma in the frame of the fluid $E_{j}^{\m{fluid}}$ by

\be
E_{j}^{\mathrm{fluid}}   = \frac{E_{j}}{\gamma_{\mathrm{bulk}}^{2}}, \qquad \frac{E_{j}}{\gamma_{\mathrm{bulk}}^{2}} = U_{e} + U_{B},
\label{Ejdef}
\ee

where $U_{e}$ and $U_{B}$ are the electron energy and magnetic field energy contained in the first metre of the jet in the $x$-direction, in the fluid frame.  We define the equipartition fraction $A_{\mathrm{equi}}$ as

\be
A_{\mathrm{equi}} = \frac{U_{B}}{U_{e}}.
\ee

Let us now consider the electron population in more detail.  It is believed from optically thin synchrotron observations of AGN that the electron energy distribution is described by a single power law.  We impose an exponential cutoff after $E_{\m{max}}$ i.e.

\begin{equation}
N_{e}(E_{e})=AE_{e}^{-\alpha}e^{-\frac{E_{e}}{E_{\rm max}}},
\end{equation}

where $N_{e}(E_{e})$ is the initial injected electron distribution contained in the first metre of the jet, $A$ and $\alpha$ are constants and $\alpha$ is thought to be between 1 and 3 from the theory of shock acceleration (\cite{1978MNRAS.182..147B}, \cite{2011MNRAS.tmp.1506B} and \cite{2011arXiv1110.5968S}).  We assume that the electron population within a section of length ${\rm d}x$ is independent of $x$. 

We define $B_{0}$ to be the magnetic field strength at the base of the jet in the lab frame, at the base of the jet $U_{e}$ and $U_{B}$ are given by

\be
U_{e}= \int_{E_{\m{min}}}^{E_{\m{max}}} E_{e} N_{e}(E_{e}) \rmn{d}E_{e}, \qquad U_{B} = \pi R_{0}^{2} \frac{B_{0}^{2}}{2\mu_{0}},
\ee

\be
U_{e}=\frac{E_{J}}{\gamma_{\mathrm{bulk}}^{2}(1+A_{\mathrm{equi}})}, \qquad U_{B} = \frac{A_{\mathrm{equi}} E_{j}}{\gamma_{\mathrm{bulk}}^{2}(A_{\mathrm{equi}}+1)}.
\ee

We can use these equations to calculate the coefficient A and to find a relation between $B_{0}$ and $R_{0}$.

\be
A=\frac{(2-\alpha)U_{e}}{(E_{\mathrm{max}}^{2-\alpha}-E_{\mathrm{min}}^{2-\alpha})} = \frac{(2-\alpha)E_{j}}{\gamma_{\mathrm{bulk}}^{2}(1+A_{\mathrm{equi}})(E_{\mathrm{max}}^{2-\alpha}-E_{\mathrm{min}}^{2-\alpha})},
\ee

\be
R_{0}=\sqrt{\frac{2U_{B}\mu_{0}}{ \pi l_{c}B_{0}^{2}}}= \sqrt{\frac{2E_{j}A_{\mathrm{equi}}\mu_{0}}{\gamma_{\mathrm{bulk}}^{2}( \pi B_{0}^{2})(1+A_{\mathrm{equi}})}}.
\ee

We choose $B_{0}$ to be our independent parameter and determine $R_{0}$ from the above equation.  For reasonable values of $B_{0}\leq 10^{-4}\m{T}$, $W_{j}>10^{34}W$ and $\gamma_{\mathrm{bulk}}<100$ (\cite{1998ApJ...509..608T} and \cite{2010arXiv1010.2856L}), we find $R_{0}>3 \times10^{11}\m{m}$.

\subsection{Magnetic Field Strength}

For a jet with a constant bulk Lorentz factor which conserves magnetic energy in each segment the magnetic field will change as a function of the radius of the jet so that the total magnetic energy is conserved in a segment.

\be
U_{B}=\pi R(x)^{2} \frac{B(x)^{2}}{2\mu_{0}}, \qquad B(x)=\sqrt{\frac{2 \mu_{0} U_{B}}{\pi R(x)^{2} }}.
\label{UB}
\ee

Substituting in the value of $U_{B}$ from equation $\ref{UB}$ we find.

\be
B(x) = B_{0} \frac{R_{0}}{R(x)}.
\ee

In this case the local magnetic energy is conserved as well as the radial and toroidal magnetic flux, however, the magnetic flux parallel to the jet axis increases with $x$.  If the magnetic field is randomly oriented on small scales then the total flux parallel to the jet axis averages to zero, so the magnetic flux parallel to the jet axis does not change along the jet.

\section{Electron radiative losses}

In our calculations we evolve the electron population dynamically along the jet by taking into account energy losses from synchrotron and inverse-Compton radiation.  For simplicity we will consider the electrons in a slab of width one metre in the fluid frame as it propagates along the jet (or more correctly, since the slab is at rest in the fluid frame, the jet structure moves relative to the slab).  As a section of the jet of width $\m{d}x$ in the fluid frame moves past an expanding stationary slab, the electrons in the slab loose an amount of energy equal to the total power emitted by the entire section in the fluid frame divided by $c$, independently of the width $\m{d}x$ of the section.  This is because at any point the section contains $n=\m{d}x$ one metre slabs and the end of the section (which is moving in the fluid frame) travels with speed c towards the slabs, so a slab takes $\frac{n}{c}$ seconds to cross the section in the fluid frame.  The evolution of the electron population along the jet due to energy losses can then be calculated using the equation below.

\be
N_{e}(E_{e},x+\m{d}x)=N_{e}(E_{e},x)-\frac{P_{tot}(x,\m{d}x,E_{e})}{cE_{e}}, \label{loss}
\ee

where $P_{tot}$ is the total power emitted by electrons of energy $E_{e}$ by a section of jet of width $\m{d}x$ in the fluid frame due to synchrotron and inverse-Compton processes.  This equation is equivalent to solving the Fokker-Planck equation for an electron distribution contained in a section of plasma which loses energy radiatively whilst moving along the jet.  We can take into account the dynamic electron population by modifying the variable $A$ to include a dependence on jet length and electron energy $A(x,E_{e})$, which takes into account the changes to the electron population as a function of jet length and electron energy. 

\be
A(x,E_{e})=A\times \frac{N_{e}(E_{e},x)}{N_{e}(E_{e},x=0)}.
\ee

We can now calculate the synchrotron and inverse-Compton emission of a variable electron population given by

\be
N_{e}(E_{e},x)=A(E_{e},x)E_{e}^{-\alpha} \times e^{-\frac{E_{e}}{E_{\m{max}}}}.
\ee

\section{Expansion losses}

In this work we use a uniform conical jet model with a constant opening angle and bulk Lorentz factor.  This model represents an overpressured jet which is expanding ballistically.  An overpressured jet with a relativistic equation of state ($p \propto \rho^{\frac{4}{3}}$) expands freely with a constant opening angle and its expansion is essentially ballistic after a few scale heights (\cite{1984RvMP...56..255B}).  A uniform conical jet with a constant bulk Lorentz factor is necessarily ballistic.  This is because its radial velocity profile is $v(r)=v_{0}\frac{r}{r_{\m{max}}}$, so the kinetic energy of plasma associated with the radial expansion is constant along the jet, also, since the conical jet occupies a fixed volume in space in the lab frame it does no work expanding its environment.  This means that the expansion does not result in adiabatic losses in this ballistic model and so there is no conversion of the internal energy of the plasma into bulk kinetic energy.  

We will investigate the effect of including adiabatic losses on a jet in Paper II.

\section{Synchrotron emission}

We use the expression for the power emitted by an electron whose velocity is at an angle $\theta$ to a uniform magnetic field of strength $B(x)$ (see for example \cite{1970ranp.book.....P})

\begin{equation}
P(E_{e},\theta,x)=\frac{e^{4}\gamma^{2}B(x)^{2}\beta^{2}\sin^{2}\theta}{6\pi\epsilon_{0}cm_{e}^{2}}.
\end{equation}

For high energy electrons, synchrotron radiation is beamed along the velocity vector of the electron.  In our model electrons are isotropically distributed in pitch angle, so integrating over $\theta$ we find

\begin{equation}
P_{\mathrm{single}}(E_{e},x) = \frac{2}{3\mu_{0}}\sigma_{T}cB(x)^{2}\beta^{2}\gamma^{2}.
\end{equation}

Considering beaming effects and rotation frequency we make the simplifying assumption that electrons emit all their energy at a critical frequency $\nu_{c}$ (this is a reasonable assumption if we integrate over a large range of electron energies) which is given by

\begin{equation}
\nu_{c}(x) = \frac{3\gamma^{2}eB(x)}{4\pi m_{e}},
\end{equation}

see \cite{1994hea..book.....L}.  Rearranging the above equation we can define the useful relations

\begin{equation}
E_{e}=(\epsilon(x) \nu_{c}(x))^{0.5}, \qquad \epsilon(x)=\frac{4 \pi m_{e}^{3}c^{4}}{3eB(x)}.
\end{equation}

We want to find the synchrotron emission from a slab in the jet at a distance $x$ with width $\m{d}x$.  The total power emitted at a critical frequency is simply the number of electrons which emit at that frequency multiplied by the power they emit.

\begin{eqnarray}
P_{\mathrm{tot}}(\nu,x) &=& P_{\mathrm{single}}(\nu,x)N_{e}(\nu,x) \rmn{d}x,\\
P_{\mathrm{tot}}(\nu,x) &=& \dfrac{\sigma_{T}B^{2}\beta^{2}A(E_{e},x)\epsilon(x)(\epsilon(x) \nu)^{\frac{1-\alpha}{2}}\rmn{d}x}{3\mu_{0}m_{e}^{2}c^{3}}.
\end{eqnarray}

This gives our total emitted power as a function of frequency in the centre of momentum frame of the electrons.  

\subsection{Opacity}

To calculate the observed emission through the jet fluid we need to know the opacity of the plasma.  We calculate the opacity using the one dimensional radiative transfer equation.  Neglecting scattering in a homogeneous plasma we find (see \cite{1991bja..book.....H})

\begin{equation}
I_{\nu} = \frac{j_{\nu}}{k_{\nu}}(1-e^{-k_{\nu}D}),
\label{radtransf}
\end{equation}

where $I_{\nu}$ is the emitted flux, $j_{\nu}$ is the emissivity per unit volume, $k_{\nu}$ is the opacity and $D$ is the length of  a section of jet plasma along our line of sight.  Using the Rayleigh-Jeans limit and associating
the thermal temperature of electrons emitting at their critical frequency with the kinetic energy of those electrons via $k_{b}T=E_{e}=(\epsilon \nu)^{0.5}$ we obtain 

\be
I_{\nu} = \frac{2 \epsilon^{\frac{1}{2}}\nu^{\frac{5}{2}}}{c^{2}} = \frac{j_{\nu}}{k_{\nu}}, \qquad k_{\nu}(\nu,x) = \frac{j_{\nu}c^{2}}{2\epsilon^{\frac{1}{2}}\nu^{\frac{5}{2}}}.
\ee

To simplify the geometry of the problem we assume that we observe at an angle smaller than the opening angle of the jet $\theta_{\m{observe}}$ in the lab frame.  We consider the emission from sections of the jet of width ${\rm d}x$, where the opacity is a function of $x$.  

\begin{eqnarray}
j_{\nu}(\nu,x) &=& n_{e}(\nu,x)P(\nu,x), \\
j_{\nu}(\nu,x) &=& \frac{2A(E_{e},x)\epsilon\sigma_{T}B^{2}\beta^{2}(\epsilon \nu)^{\frac{1-\alpha}{2}}}{3\pi R(x)^{2} \rmn{d}x \mu_{0} m_{e}^{2}c^{3}} \rmn{d}x,\\
j_{\nu}(\nu,x) &=& j_{0}(E_{e},x)\nu^{\frac{1-\alpha}{2}}.
\end{eqnarray}

So using this in our expression for opacity we find

\begin{equation}
k_{\nu}(\nu,x) = \frac{j_{0}(E_{e},x)c^{2}\nu^{-\frac{\alpha+4}{2}}}{2\epsilon^{0.5}}.
\end{equation}

Using Equation $\ref{radtransf}$ we see that if $k_{\nu}D$ is small (the material is optically thin) the observed luminosity simplifies to the volume of the plasma multiplied by the emissivity as we would expect.  If we observe down the jet within the opening angle we will see through to different depths at different wavelengths due to the wavelength dependence of the opacity.  We see through to the layer where the total column optical depth is approximately 1 and beyond this there is an exponentially decreasing contribution from further layers.  The total optical depth $\tau_{tot}$ is the sum of the optical depths from each section. 

In order to calculate the optical depth of a section with lab frame width $\m{d}x'$, for a photon travelling at an angle $\theta_{\m{observe}}$ to the jet axis in the lab frame, we need to Lorentz transform its 4-displacement from the lab to the fluid frame, since we have calculated the synchrotron opacity in the fluid frame.  We require the distance the photon travels in the fluid frame $\m{d}r$, this is related to the distance travelled by the photon in the lab frame by a simple Lorentz transformation.  The 4-displacement in the lab frame is given by

\be
X'=\begin{pmatrix}\, c\, \m{d}t'\, \\ \m{d}x' \\ \m{d}y'  \end{pmatrix}=\begin{pmatrix} \frac{\m{d}x'}{\cos(\theta_{\m{observe}})} \\ \m{d}x' \\ \m{d}x' \tan(\theta_{\m{observe}}) \end{pmatrix}.
\ee

We Lorentz transform this to the plasma rest frame to find the distance $\m{d}r$

\be
X=\begin{pmatrix} \gamma_{\m{bulk}} \m{d}x'\left(\frac{1}{\cos(\theta_{\m{observe}})}-\beta\right) \\ \gamma_{\m{bulk}} \m{d}x'\left(1-\frac{\beta}{\cos(\theta_{\m{observe}})}\right) \\ \m{d}x' \tan(\theta_{\m{observe}}) \end{pmatrix},
\ee
\be
\m{d}r=\gamma_{\m{bulk}} \m{d}x' \left(\frac{1}{\cos(\theta_{\m{observe}})}-\beta\right).
\label{dr}
\ee

For a section of lab frame width $\m{d}x'$ the optical depth is simply the fluid frame synchrotron opacity multiplied by the distance $\m{d}r$

\be
\tau_{tot}(\nu,x) = \int_{x}^{L} k_{\nu}(\nu,x) \gamma_{\m{bulk}}^{2} \left(\frac{1}{\cos(\theta_{\m{observe}})}-\beta \right) \rmn{d}x,
\ee

where we have converted the integration over $\m{d}r$ to $\m{d}x$ using Equation $\ref{dr}$ and the relation $\m{d}x'=\gamma_{\m{bulk}}\m{d}x$ (due to Lorentz contraction). An individual segment of width ${\rm d}x$ will emit according to

\be
P_{\nu}(x,\rmn{d}x,\nu)=\pi R(x) \m{d}x \frac{2 \epsilon^{\frac{1}{2}}\nu^{\frac{5}{2}}}{c^{2}}(1-e^{-k_{\nu}(\nu,x) R(x)}).
\label{synchloss}
\ee

We use the equation above to calculate the electron energy losses via Equation $\ref{loss}$ since this is the expression for the total synchrotron power emitted by a segment of the jet.  We will observe the emission from each segment weighted by the total optical depth, which is equivalent to solving the radiative transfer equation with $j_{\nu}(z)=0$.

\be
P_{\nu}= \sum_{x} P_{\nu}(x,dx,\nu) e^{-\tau_{tot}(\nu,x)} . \label{synch1}
\ee

The total synchrotron emission in the fluid frame, $P_{\nu}$, is the sum of the emission from all the sections in the jet multiplied by fraction of radiation absorbed along the jet from a section.

\section{Inverse-Compton Scattering}

In this section we calculate the inverse-Compton emission from an arbitrary electron population and arbitrary isotropic seed photon population.  Let us consider the most general case of an electron-photon collision.  In the lab frame we have an electron with energy $E_{e}$ travelling in a direction, we are free to define our $x$-axis (not to be confused with the jet $x$-axis) to be aligned with the electron's momentum by a simple rotation.  In these lab coordinates we have an incoming photon with energy $E_{\gamma}$ with its momentum vector at an angle of $\theta$ to the negative $x$ direction.  We wish to calculate the scattered photon energy and angle in the lab frame after the collision.

\begin{figure}
\centering
\includegraphics[width=8cm]{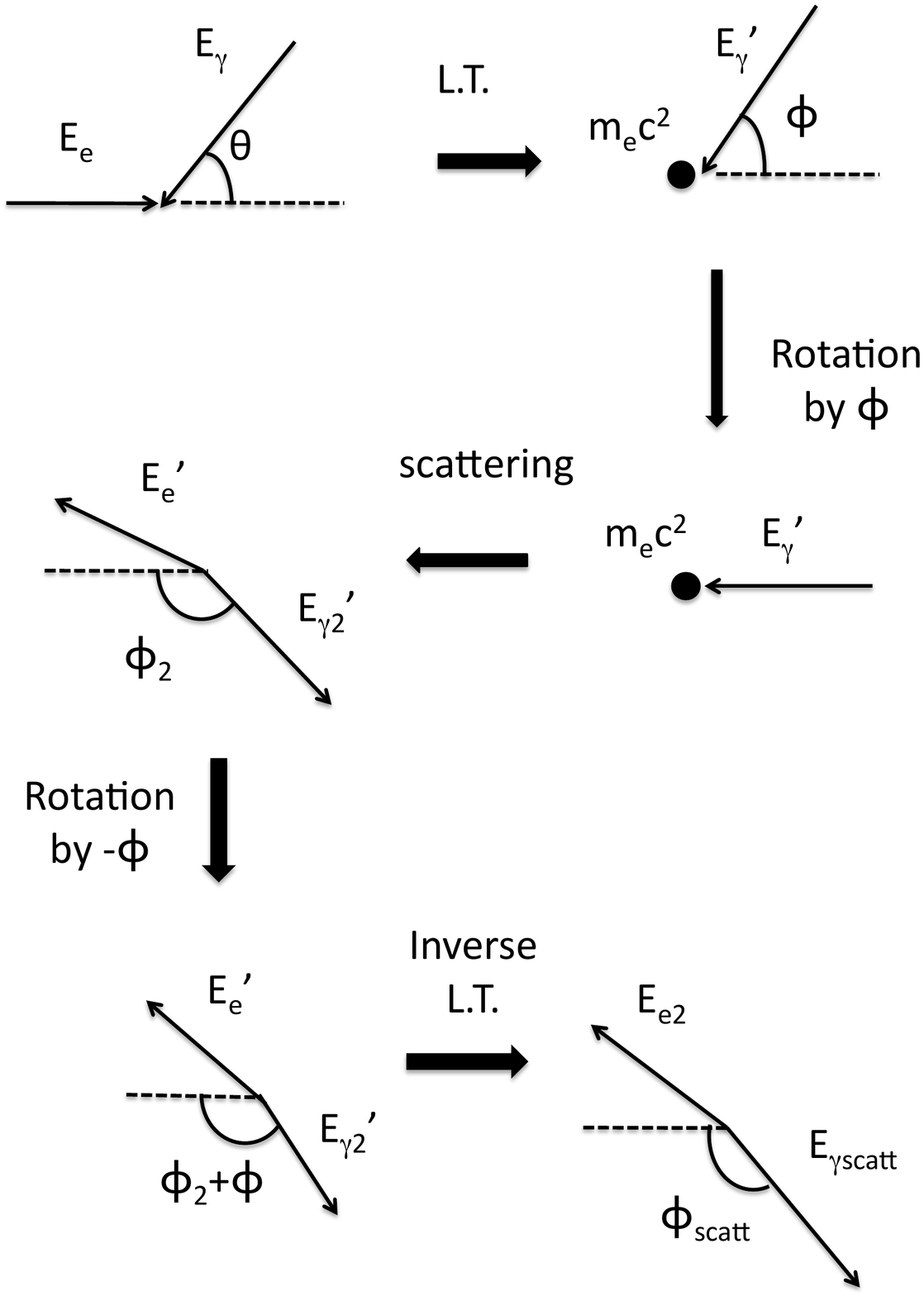}
\caption{Schematic diagram of the inverse-Compton scattering process defining our choice of variables.}
\label{ICscatt}
\end{figure}

To simplify the calculation we Lorentz transform into the electron's rest frame (where the Klein-Nishina cross section is well defined) rotate our coordinates to deal with a simple stationary head-on collision, then apply a further rotation to realign our coordinate system with our original lab $x$-axis and finally perform the inverse Lorentz transformation to revert to our initial lab frame.  This is illustrated schematically in Figure $\ref{ICscatt}$ and is calculated using standard four-vector Lorentz transforms. 

\subsection{Obtaining the Inverse-Compton Spectrum}

We have calculated the scattering angle and energy of a photon which is scattered by an electron.  It remains to relate this to an emitted flux.  We approach this problem by discretely summing over interval values of all the dynamic variables ($E_{e}$, $E_{\gamma}$, $\theta$ and $\phi_{2}$ and $x$) and assigning a weight to each of the scattered photons which corresponds to the number of photons produced per second in a section of width ${\rm d}x$ in the fluid frame.  We then sum over the contributions from each section of the jet.

\bea
\mathrm{Power} = \int_{E_{e}, E_{\gamma}, \theta, \phi_{2},x} E_{\gamma \mathrm{scatt}}.\mathrm{weight}(E_{e}, E_{\gamma}, \theta, \phi_{2},x) \nonumber \\
\times \rmn{d}E_{e} \rmn{d}E_{\gamma} \rmn{d}\theta \rmn{d}\phi_{2} \rmn{d}x.
\label{Power}
\eea

\subsection{Klein-Nishina Cross Section}

To calculate the weight function we need the cross section for the electron-photon interaction.  The Klein-Nishina cross section is the first order QED cross section for electron photon scattering (see for example \cite{1995qtf..book.....W}), it is given by

\begin{equation}
\frac{\rmn{d}\sigma}{\rmn{d}\Omega_{2}} = \frac{1}{2} \alpha_{\m{FS}}^{2} r_{c}^{2} P(E_{\gamma}', \phi_{2})^{2}(P(E_{\gamma}', \phi_{2})+P(E_{\gamma}', \phi_{2})^{-1}-1+\cos^{2}\phi_{2}),
\end{equation}

\begin{equation}
r_{c}= \frac{\hbar}{m_{e}c}, \qquad P(E_{\gamma}', \phi_{2})= \frac{1}{1+\frac{E_{\gamma}'}{m_{e}c^{2}}(1+\cos{\phi_{2}})}, \nonumber
\end{equation}
\be
\rmn{d}\Omega_{2}=\pi |\sin \phi_{2}| \rmn{d}\phi_{2}.
\ee

where $\alpha_{\m{FS}}$ is the fine structure constant.  The formula relating the number of collisions between electrons of energy $E_{e}$ and photons of energy $E_{\gamma}$ is 

\bea
\mathrm{Collisions}(\mathrm{s}^{-1}) = \int N_{e}(E_{e},x).\frac{\rmn{d}\sigma(E_{\gamma}', \phi_{2})}{\rmn{d} \Omega_{2}}.n_{\gamma}(E_{\gamma}, \theta,x) \nonumber \\
\times c(1-\beta \cos\theta).\rmn{d}\Omega_{2} \rmn{d}E_{e} \rmn{d}E_{\gamma} \rmn{d}\Omega \rmn{d}x,
\eea

\begin{equation}
\rmn{d}\Omega = \pi |\sin\theta| \rmn{d}\theta,
\end{equation}

where $c(1-\beta\cos(\theta))$ is the relative velocity of the photon and electron (see for example \cite{1970RvMP...42..237B}).  Since the integral of the weight function is simply the number of collisions occurring per second we find

\bea
&&\mathrm{weight}(E_{e}, E_{\gamma}, \theta, \phi_{2},x) = N_{e}(E_{e},x).\frac{\rmn{d}\sigma(E_{\gamma}', \phi_{2})}{\rmn{d}\Omega_{2}} \nonumber \\
&& \qquad \, \times \frac{n_{\gamma}(E_{\gamma}, \theta, x)}{4 \pi}. c(1-\beta \cos\theta) \pi |\sin \phi_{2}| \pi |\sin \theta|.
\eea

We want to replace the integral in Equation $\ref{Power}$ with a discrete sum over the variables.  Under the assumption that the weighting function is approximately constant for sufficiently small  parameter intervals in the sum, we can convert the integral over the weighting function into a discrete sum 

\bea
\mathrm{Power}(W) = \sum_{E_{e}, E_{\gamma}, \theta, \phi_{2},x} E_{\gamma \mathrm{scatt}}.\mathrm{weight}(E_{e}, E_{\gamma}, \theta, \phi_{2},x) \nonumber \\
\times \rmn{d}E_{e} \rmn{d}E_{\gamma} \rmn{d}\theta \rmn{d}\phi_{2} \rmn{d}x.
\label{PowIC}
\eea

This is our expression for the emitted power in the fluid frame.  Using this method to calculate the emitted power has the advantage of calculating the inverse-Compton scattering exactly in the limit of small step sizes in the sum, whilst also obtaining the angular dependence of the spectral energy distribution.  To do this we simply carry out the sum in equation $\ref{PowIC}$ binning scattered photons depending on both their scattered angle, which can easily be related to their observation angle, and their frequency.  In this way we can calculate the emitted spectrum as a function of observation angle.

In the next section we will calculate the photon number density $n_{\gamma}$ from the synchrotron emission.

\section{Synchrotron Seed Photons}

At high observed photon energies($>1\rm{keV}$), where inverse-Compton scattering starts to dominate the blazar spectrum, the jet is optically thin to photons.  To calculate the inverse-Compton emission from segments of the jet we need to calculate the photon number density $n_{\gamma}(x,\nu)$.  When a section of the jet is optically thin ($K_{\nu}R(x)>1$) an infinitesimal section of width ${\rm d}x$ will receive approximately the same amount of radiation from neighbouring sections as it emits parallel to the $x$-axis.  We use the approximation that only the radiation emitted radially is unbalanced.  So the volume of depth one light-second from the outer radial surface contains approximately one second of emitted power.  If the section is optically thick then the emission and absorption per unit length are equal and so the photon distribution in the object is that of a blackbody.  Calculating $n_{\gamma}$ for the optically thin case we find

\be
n_{\gamma \, \nu}(x,\rmn{d}x) = \frac{P_{\nu}(x,\rmn{d}x,\nu)}{ 2 \pi h \nu R(x) \rmn{d}x},
\ee

and for the optically thick case ($k_{\nu}R(x)>1$)

\be
n_{\gamma \, \nu}(x,\rmn{d}x) = \frac{8 \pi \nu^{2}}{c^{3}(e^{\frac{h\nu}{k_{b}T}}-1)}.
\ee

We define the average photon number density $n_{\gamma}^{av}(\nu)$ 

\be
n_{\gamma}^{av}(\nu)=\sum_{x} \frac{n_{\gamma}(\nu,x) \rmn{d}x}{L},
\label{ngammaav}
\ee

\begin{figure*}
	\centering
		\subfloat[A pure SSC fit to the simultaneous 2008-9 observations.]{ \includegraphics[width=12 cm, clip=true, trim=1cm 1cm 1cm 1cm]{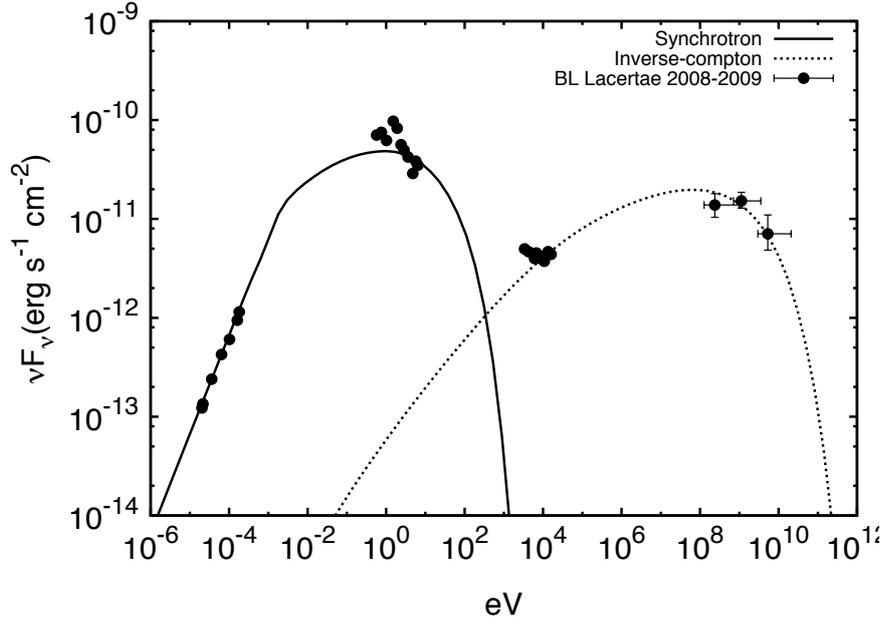} }\\
		\subfloat[This figure shows a comparison between models including and neglecting electron energy losses]{ \includegraphics[width=8 cm, clip=true, trim=1cm 1cm 1cm 1cm]{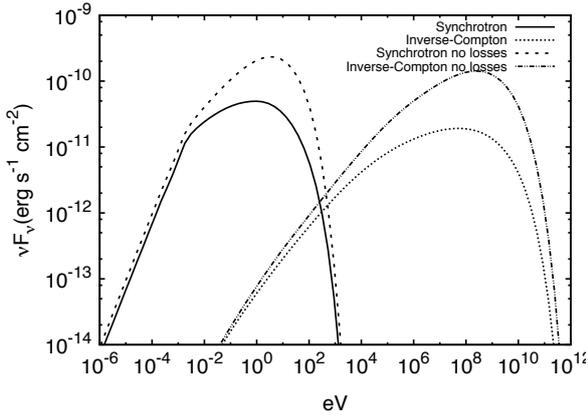} }
		\qquad
		\subfloat[The electron energy distribution contained in a slab of fixed width one light-second in the fluid frame, shown after injection at the base and after losses at the end of the jet.]{ \includegraphics[width=8cm, clip=true, trim=1cm 1cm 1cm 1cm]{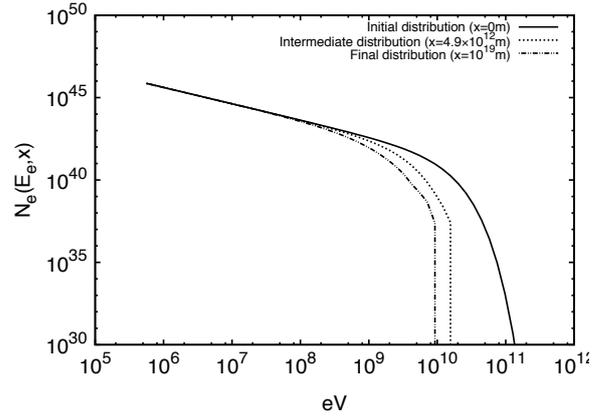} }
	\caption{\emph{Top:} The observed spectrum of the model fitted to the recent simultaneous multi-wavelength observations of BL Lacertae.  The model reproduces the observed spectrum at all frequencies \emph{Bottom:} These figures show the effect of electron energy losses to the jet.  The figure on the right shows that the high energy electrons have been depleted by energy losses at the end of the jet.  The figure on the left shows that energy losses have the largest effect on the high frequency synchrotron and inverse-Compton emission of the jet, but do not affect the flat radio spectrum.}
	\label{fig:1}
\end{figure*}

\section{Doppler Boosting of Emission}

So far we have calculated synchrotron and inverse-Compton emission in the fluid frame.  In our model the fluid is moving with a bulk Lorentz factor $\gamma_{\mathrm{bulk}}$ along the jet axis so the emission will be boosted depending on the Lorentz factor and the observation angle to the jet $\theta_{\m{observe}}$.  We use primed variables to denote quantities defined in the lab frame and unprimed variables for quantities defined in the fluid frame.  The lab frame jet opening angle $\theta_{\m{opening}}'$ is related to the fluid frame opening angle $\theta_{\m{opening}}$ via

\be
\gamma_{\m{bulk}} \tan \theta_{\m{opening}}' = \tan \theta_{\m{opening}},
\ee

due to length contraction of the jet axis of the cone.  The observed photon frequency in the lab frame is boosted by a Doppler factor (shown below) which is obtained by Lorentz transforming a photon from the jet fluid frame into the lab frame.  The total observed power is related the emitted power in the fluid frame by multiplying by four Doppler factors one from Doppler shift in frequency, one from increased photon arrival times and two from transforming solid angle between frames.

\begin{equation}
\frac{1}{\gamma(1-\beta \cos \theta_{\m{observe}})}.
\end{equation}

The observed power in solid angle $d\Omega'$ at a Doppler boosted frequency $\nu' $, $P'(\nu')$, can then be related to the synchrotron power emitted in the fluid frame at frequency $\nu$ emitted into solid angle $\rm{d}\Omega$, $P(\nu)$, via

\begin{eqnarray}
P'(\nu' ) &=& P'\left(\nu \frac{1}{\gamma(1-\beta \cos\theta_{\m{observe}})}\right) \\&=&  P(\nu) \left[\frac{1}{\gamma (1-\beta\cos \theta_{\m{observe}})}\right]^{4} .
\end{eqnarray}

BL Lacertae has the very small cosmological redshift of $z=0.0686$, so we take the distance to BL Lacertae be $d=300\rm{Mpc}$ and assume Euclidean space.

\section{Results}

\subsection{Modelling the SED of BL Lacertae}

In Figure $\ref{fig:1}$a we show the results of our fluid jet model fitted by eye to the recent simultaneous muti-wavelength observations of BL Lacertae ( \cite{2011ApJ...730..101A}).  We stress how well our relatively simple model fits these observations:  the model fits well to the data across almost all wavelengths using only inverse-Compton scattered synchrotron seed photons from the single population of jet electrons. There is a discrepancy at the lower end of the x-ray energies, and we argue that this is likely from the accretion disc corona (see, e.g., \cite{2011arXiv1109.2069J}).
The optical/IR data is clearly not consistent with a smooth synchrotron component, but has been necessarily obtained with multiple instruments and may therefore suffer from absolute calibration offsets. By contrast, the optical-near IR spectrum on BL Lac obtained by \cite{1995ApJ...452L...5V} shows that the spectrum is clean and continuous in this region. 

Figures $\ref{fig:1}$b and $\ref{fig:1}$c show the effect of including electron energy losses in the model.  In Figure $\ref{fig:1}$c we can see the evolution of the electron energy distribution along the jet.  The figure shows the electron energy spectrum (contained in a slab of width one light-second) at the base of the jet immediately after injection and at the end of the jet after the slab has propagated through the entire jet and suffered maximal energy losses.  We can see that losses are most significant for the highest energy electrons as we would expect, since the power of both synchrotron and inverse-Compton emission is approximately proportional to $\gamma^{2}$ of the electrons.  The energy losses effectively lower the maximum energy cutoff of the electrons as they move along the jet.  

Energy losses are most severe close to the base of the jet due to its stronger magnetic field and smaller radius.  This means that losses occur predominantly in a region close to the base which is much smaller than the total jet length.  This results in the jet producing a flat radio spectrum even when taking into account radiative energy losses since the electron distribution does not change significantly along the vast majority of the jet.  We find that the base of the jet dominates the synchrotron and inverse-Compton emission at the highest frequencies whilst the radio synchrotron emission, which is optically thick at the base, is dominated by the outer regions of the jet with larger radii and therefore a larger surface area for emission. 

In Figure $\ref{fig:1}$b we can see the effect of the energy losses on the emitted spectrum of the jet.  The figure shows that the low frequency synchrotron emission does not change significantly when losses are taken into account, however, the optically thin synchrotron emission which is dominated by the highest energy electrons at the base of the jet is reduced.  The effect of losses on the highest energy electrons also results in a reduction to the highest frequency inverse-Compton emission, for which they are responsible.

\begin{table}
\centering
\begin{tabular}{| c | c |}
\hline
Parameter & Value \\ \hline
$W_{j}$ & $7.63 \times 10^{36}\rm{W}$ \\ \hline
L & $1 \times 10^{19}\rm{m}$ \\ \hline
$B_{0}$ & $3.63 \times 10^{-5} \rm{T}$ \\ \hline
$R_{0}$ & $7.32 \times 10^{13}\rm{m}$ \\ \hline
$A_{\m{equi}}$ & 1 \\ \hline
$E_{min}$ & 5.11 \rm{MeV} \\ \hline
$E_{max}$ & 5.60 \rm{GeV} \\ \hline
$\alpha$ & 2 \\ \hline
$\theta'_{\m{opening}}$ & $9.7^{o}$ \\ \hline
$\theta_{\m{observe}}$ & $2^{o}$ \\ \hline
$\gamma_{\m{bulk}}$ & 12 \\ 
\hline
\end{tabular}
\caption{This table shows the values of the physical parameters used in the model used to fit to the BL Lacertae observations}
\end{table}

From fitting to the observations of BL Lacertae we find a jet power of $7.63\times 10^{36}\rm{W}$, a maximum magnetic field $B_{0}=3.63 \times 10^{-5}\rm{T}$ at the base of the jet with base radius $R_{0}=7.32 \times 10^{13}\rm{m}$ and lab frame jet length $1.2 \times 10^{20}\rm{m}$.  We also find a bulk Lorentz factor $\gamma_{\m{bulk}}=12$, an electron energy index $\alpha=2.0$  and an observation angle $\theta_{\m{observe}}=2^{o}$.  We have fitted the observations with an equipartition jet $A_{\m{equi}}=1$ corresponding to energy being equally split between the electron population and the magnetic field.  We have chosen the jet to be in equipartition because this is compatible with observations of radio lobes of Seyfert 1 galaxies being approximately in equipartition (\cite{2007ApJ...668..203M}). 

In this model we have not included any reacceleration of the electrons through internal shocks in the jet.  Internal shocks are a mechanism which can convert some of the large store of kinetic energy in the bulk motion of the jet into internal energy in the electron population.  This has been investigated by a number of authors (\cite{1985ApJ...298..114M}, \cite{1998A&A...333..452K}, \cite{2000A&A...356..975K}, \cite{2001MNRAS.325.1559S}, \cite{2004ApJ...613..725S} and \cite{2010MNRAS.401..394J}) as a way of replenishing radiative and adiabatic energy losses of the electron population in jets.  We expect that our estimates of jet opening angle and bulk Lorentz factor are larger than if we had included a form of in situ reacceleration, since a larger jet opening angle and Doppler factor reduce the effect of energy losses on the electron population.  We intend to investigate the effect of different forms of reacceleration in this model in future work.   

The paper presenting the observations of BL Lacertae ( \cite{2011ApJ...730..101A}) uses three models to fit to the most recent observations.  The first model from  \cite{2008ApJ...686..181F} is a single-zone time independent spherical SSC (synchrotron self-Compton) model.  The fitted model uses a large Doppler factor of 26 and is over an order of magnitude from equipartition, it also fails to reproduce the low frequency radio emission.  The second model based on work by  \cite{2008bves.confE..73C} is a time-dependent single-zone SSC it uses two different spherical blobs to fit to high and low frequency synchrotron emission and it fits well to the observations.  The two blobs have Doppler factors 10 and 6.5 and the average equipartition value is $A_{\m{equi}}=2.8$.  The final model used is based on \cite{2002ApJ...581..127B} and is a single-zone time-dependent model with ERC (external radiation Compton) and also fits well to the data.  The fit has a Doppler factor of 15, an equipartition fraction $A_{\m{equi}}=1.48$ but also includes an energy independent particle escape time of $t_{\mathrm{esc}}=60R/c$, which is difficult to justify physically.  Whilst these models are able to fit the high energy synchrotron and inverse-Compton emission they are not able to reproduce the observed flat radio slope of the spectrum, the self-absorbed spectra they produce are steeper.     

One of the major achievements of the conical jet model is to reproduce the radio emission observed at low frequencies.  This cannot be reproduced in fixed radius, homogeneous, one-zone SSC, two-zone SSC and ERC models (\cite{2010ApJ...716...30A} and \cite{2003ApJ...593..667M}) without introducing additional complications such as a two power law fit to the electron population or energy-independent diffusion of electrons.  In fitting to the radio observations as well as higher frequencies we gain additional constraints on our model parameters.  In general we can gain the electron energy spectral index from the slope of the inverse-Compton emission (see for example \cite{1994hea..book.....L}), the turnover of the flat spectrum at high frequencies then depends on the magnetic field strength $B_{0}$ which is strongly constrained by the synchrotron turnover at $\sim 1 \rm{keV}$ and spectral index.  Since blazars have such a variety of spectral behaviour with increasing variability at higher frequencies (see \cite{1997ARA&A..35..445U}) it is surprising that most models neglect reproducing observations in the radio, since for most blazars the radio is less variable and normally is observed to be close to a flat spectrum for a wide range of low frequencies.  By not fitting to radio observations, models reduce the number of constraints and lack a physical mechanism to reproduce one of the most stable features of blazars. 

Whilst our model has a number of free parameters we find in fitting the model to the observations that there are not large degeneracies between the different physical parameters.  This is because the different parameters affect specific aspects of the emission which are well constrained by the multi-wavelength observations.  We find that increasing the jet power increases the emission across all frequencies without altering the shape of the spectrum significantly.  The dependence of the inverse-Compton emission on both the synchrotron power emitted (SSC photons) and the total electron kinetic energy, which both depend on the jet power, leads to the inverse-Compton emission depending more sensitively on the jet power than the synchrotron emission.  The electron spectral index $\alpha$ determines the slope of the optically thin synchrotron emission and the slope of the inverse-Compton emission which is reasonably well constrained by the data.  The magnetic field strength at the base of the jet $B_{0}$ (which fixes $R_{0}$ if $W_{j}$ is specified) changes the frequency at which the synchrotron emission becomes optically thin and the power and peak frequency of the synchrotron emission.  The jet opening angle $\theta'_{\m{opening}}$ has a negligible effect on the high energy emission and only weakly effects the power emitted at a given radio frequency (smaller opening angles gives more radio power).  The length of the jet $L$ determines the lowest radio frequencies which are still optically thick to emission, this parameter is not well determined by the data since it has very little effect on the observable emission.  Finally, $\gamma_{\m{bulk}}$ and $\theta_{\m{observe}}$ together determine the Doppler boosting of the jet.  The observation angle determines the optical path length travelled by synchrotron photons, but this has a relatively small effect, so it primarily changes the Doppler factor for a given value of $\gamma_{\m{bulk}}$.  The bulk Lorentz factor of the jet determines the Doppler factor but also has a more subtle impact on other jet parameters such as the jet opening angle in the fluid frame and the initial energy density in plasma in the fluid frame (see Equation $\ref{Ejdef}$).  This means that the bulk Lorentz factor has a significant effect on both the observed power and spectral shape of the emission.      

\begin{figure}
	\centering
		\includegraphics[width=7 cm]{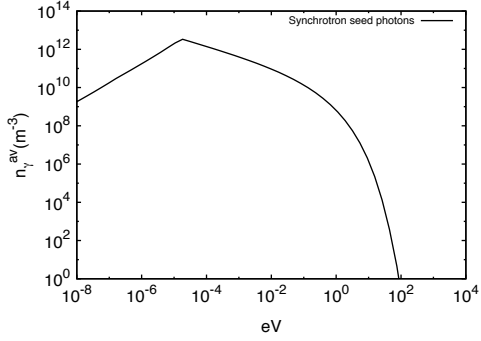} 
		
	\caption{\emph{Top:} The average number density of photons plotted against energy for the model fitted to the simultaneous BL Lac observations.}
	\label{fig:2}
\end{figure}

Figure $\ref{fig:2}$ shows the seed photon population averaged over the jet $n_{\gamma}^{av}$ (see equation $\ref{ngammaav}$) for the models in Figure $\ref{fig:1}$.  Fundamentally an ERC model with an unconstrained seed photon population can be used to fit almost any high frequency observation.  So it is important to explicitly show the seed photon population used when fitting the model to the data to ensure clarity and to confirm that the photon population is realistic.  

\begin{figure}
\centering
\includegraphics[width=8.5cm]{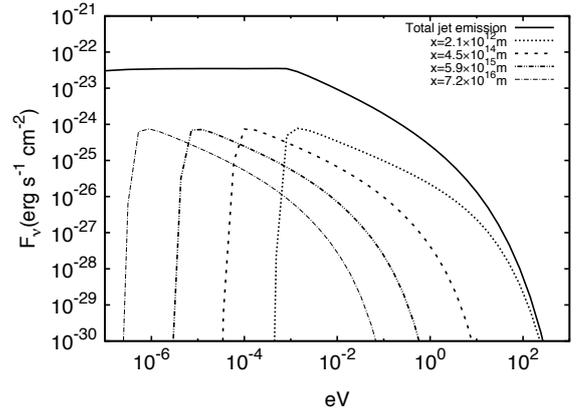}
\caption{This figure shows the observed luminosity of the BL Lacertae model looking at a small angle to the jet axis.  The solid line is the total luminosity which is flat over a wide range in photon energy.  The dashed lines show the observed contributions from thin slabs at different distances $x$ along the jet which add to produce the flat total distribution.} 
\label{fig3}
\end{figure}

\subsection{Producing a Flat Radio Spectrum}

The observation of a near flat radio spectrum is a characteristic of most blazars (\cite{2010ApJ...716...30A}).  We find that our model produces a flat spectrum which is largely insensitive to the physical parameters of the jet.

Figure $\ref{fig3}$ shows the overall flat spectrum produced by the model.  The dashed curves show the observed emission from components at different distances along the jet.  The figure illustrates the classic explanation of a flat spectrum; that different components have different turnover frequencies which add to give a flat spectrum.  We believe that this is the first time this component explanation of a flat spectrum has been calculated explicitly for a conical jet orientated close to our line of sight.  The figure shows that components closer to the base of the jet have turnovers at the highest frequencies and dominate the high frequency synchrotron emission (due to stronger magnetic fields and smaller radii) before their observed low frequency emission is absorbed exponentially by the outer jet material.  

We have found that the production of a flat radio spectrum is largely insensitive to the parameters used in our model although the magnitude and frequency range of the flat spectrum do show a dependence.  The most sensitive parameters are the overall jet length which affects the extent of the flat spectrum to low frequencies and the value of $B_{0}$ which affects the departure from the flat spectrum at high frequencies.  This means that a flat radio spectrum is intrinsic to the uniform conical jet model which conserves magnetic energy and requires no fine-tuning. 

\begin{figure}
\centering
\includegraphics[width=8.5cm]{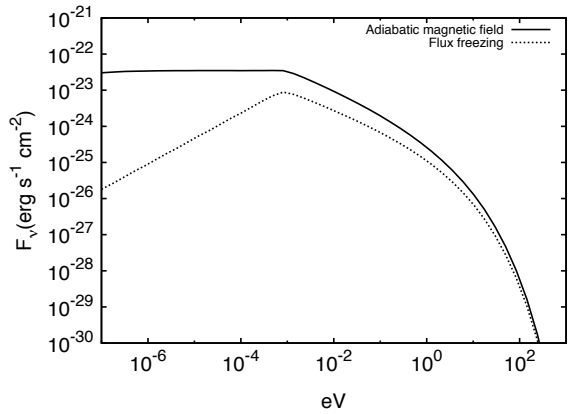}
\caption{This figure shows the dependence of the synchrotron emission on the functional form of $B(x)$.  We can clearly see that if the magnetic energy is conserved ($B(x) \propto R^{-1}$) along the jet then the synchrotron luminosity is flat over a wide range of low frequencies.  If the magnetic field evolves to conserve the flux through the surface $x=constant$ then we no longer observe a flat radio spectrum ($B(x) \propto R^{-2}$).  Since a flat or nearly-flat radio is a well known characteristic of blazar-type objects, this is evidence of the conservation of magnetic energy in simple fluid jets.}
\label{fig5}
\end{figure}

Figure $\ref{fig5}$ shows the dependence of the synchrotron emission on the functional form of $B(x)$.  We can see that if the magnetic energy is conserved along the jet ($B \propto R(x)^{-1}$) then the synchrotron luminosity is flat over a wide range of frequencies.  If the magnetic field evolves to conserve the magnetic flux parallel to the jet axis ($B(x) \propto R(x)^{-2}$) then we no longer observe a flat radio spectrum.  Since a flat or nearly-flat radio flux is a well known characteristic of blazar-type objects, this is evidence for the conservation of magnetic energy in simple fluid jets.  This result was obtained by \cite{2006MNRAS.367.1083K} who investigated a variety of fluid jet models and found in agreement with us that a model which conserved magnetic energy produced a flat spectrum whilst a model where magnetic flux is frozen did not.  However, his calculations were limited to the case of observations perpendicular to the jet axis.

Magnetic flux should be conserved along the jet both parallel and perpendicular to the jet axis.  We have found that observations seem to indicate that the magnetic energy is conserved in a section of the jet as it expands and that magnetic flux perpendicular to the jet axis is conserved, whilst flux parallel to the jet axis increases along the jet.  In order to conserve flux parallel to the jet axis this means that either the magnetic field is oriented randomly on small scales and so the net flux parallel to the jet is zero or the magnetic field in the jet is mainly toroidal and not poloidal.  This is an interesting result since dominant toroidal magnetic fields in jets are predicted by MHD simulations and theory (\cite{2005MNRAS.360..869L}, \cite{2006MNRAS.368.1561M} and \cite{2007MNRAS.380...51K}) and are supported by polarisation measurements and Faraday rotation maps (\cite{2009MNRAS.393..429O} and \cite{2009ApJ...694.1485K}).     

\section{Conclusions}

We have re-visited the uniform conical jet fluid jet model for AGN jets and for the first time used it to fit both the synchrotron and self-Compton components of the jet spectrum. In particular we have shown good agreement with these components in the recent multi-wavelength observations of the relatively low-jet-power source BL Lacertae from radio to \gray. Our model and conclusions may be summarized as: 

\begin{itemize} 

\item We demonstrate that a fluid model of an axisymmetric jet with a fixed opening angle and conserved magnetic energy fits the simultaneous observations of BL Lacertae presented by \cite{2011ApJ...730..101A}.

\item  Our model conserves energy and particle flux along the jet, this allows us to estimate physical properties of the jet and central engine such as the total jet power and magnetic field strength from fitting the model to a blazar spectrum.   

\item We evolve the electron population dynamically along the jet taking into account energy losses from synchrotron and inverse-Compton emission.  We calculate the inverse-Compton emission by numerically integrating the exact expression and we explicitly show the synchrotron seed photon population.  Unlike previous investigations, we integrate synchrotron opacity along the line of sight through the jet plasma taking into account the emission and opacity of each section of the jet.

\item We show that a conical jet model which conserves magnetic energy produces the characteristic flat radio spectrum observed in blazars in agreement with \cite{1979ApJ...232...34B} and \cite{2006MNRAS.367.1083K}.  

\item Under our conical jet assumptions, we find that for a source of this jet power ($\sim 10^{37} W$), that there is no requirement for significant re-acceleration within the jet to explain the observed spectrum. 

\item We find that the production of a flat radio spectrum does not require fine-tuning of the model parameters.  We show that a conical jet which conserves magnetic flux parallel to the jet axis does not produce a flat radio spectrum.

\end{itemize}

In this first paper in the series we have described a flexible model for a continuous jet which includes a detailed treatment of the synchrotron and inverse-Compton emission from the jet and radiative energy losses to the electrons.  For the first time we have attempted a full-spectrum fit to a blazar using a conical jet model.  We find that BL Lac can be described by a ballistic conical jet in equipartition with realistic physical parameters without a need for significant in situ acceleration or external radiation.  In the next paper  we will consider these latter effects of the model in the context of high-jet-power Compton dominant blazars.  

\section*{Acknowledgements}

WJP acknowledges an STFC research studentship. GC acknowledges support from STFC rolling grant ST/H002456/1.  We wish to thank the anonymous referee for helping to improve the clarity of this paper.  We would like to thank Lance Miller for reading a draft of the work and offering many helpful comments.  We would also like to thank Jim Hinton, Brian Reville, Tony Bell and Richard Booth for useful discussions about the model.  

\bibliographystyle{mn2e}
\bibliography{pc}
\bibdata{pc}

\label{lastpage}

\end{document}